\documentclass[english]{article}
\usepackage[T1]{fontenc}
\usepackage[utf8]{inputenc}
\usepackage{babel}
\usepackage{color}
\usepackage{graphicx}
\usepackage{multirow}
\usepackage{cite}
\usepackage{dirtree}
\usepackage{framed}
\usepackage{amsmath}
\begin{document}

\title{First-principles data set of 45,892 isolated and cation-coordinated conformers of\\20 proteinogenic amino acids}

\author{Matti Ropo\textsuperscript{1,2,3{*}}, Markus Schneider\textsuperscript{1}, Carsten Baldauf\textsuperscript{1{*}},\\Volker Blum\textsuperscript{1,4{*}}}

\maketitle

\noindent
1.\,Fritz Haber Institute of the Max Planck Society, 14195 Berlin, Germany;
2.\,Department of Physics, Tampere University of Technology, 33720 Tampere, Finland; 
3.\,COMP, Department of Applied Physics, Aalto University, 00076 Aalto, Finland; 
4.\,Department of Mechanical Engineering and Materials Science, Duke University, Durham, NC 27708, USA.\\
{*}corresponding authors:
Matti Ropo (matti.ropo@tut.fi); Carsten Baldauf \\(baldauf@fhi-berlin.mpg.de); Volker Blum (volker.blum@duke.edu).

\begin{abstract}
\noindent
We present a structural data set of the 20 proteinogenic amino acids and their amino-methylated and acetylated (capped) dipeptides. Different protonation states of the backbone (uncharged and zwitterionic) were considered for the amino acids as well as varied side chain protonation states. Furthermore, we studied amino acids and dipeptides in complex with divalent cations (Ca$^{2+}$, Ba$^{2+}$, Sr$^{2+}$, Cd$^{2+}$, Pb$^{2+}$, and Hg$^{2+}$).
The database covers the conformational hierarchies of 280 systems in a wide relative energy range of up to 4\,eV (390\,kJ/mol), summing up to an overall of 45,892 stationary points on the respective potential-energy surfaces.
All systems were calculated on equal first-principles footing, applying density-functional theory in the generalized gradient approximation corrected for long-range van der Waals interactions.
We show good agreement to available experimental data for gas-phase ion affinities.
Our curated data can be utilized, for example, for a wide comparison across chemical space of the building blocks of life, for the parametrization of protein force fields, and for the calculation of reference spectra for biophysical applications.
\end{abstract}

\newpage
\section*{Background \& Summary}

\begin{figure}
\includegraphics[width=1.2\textwidth]{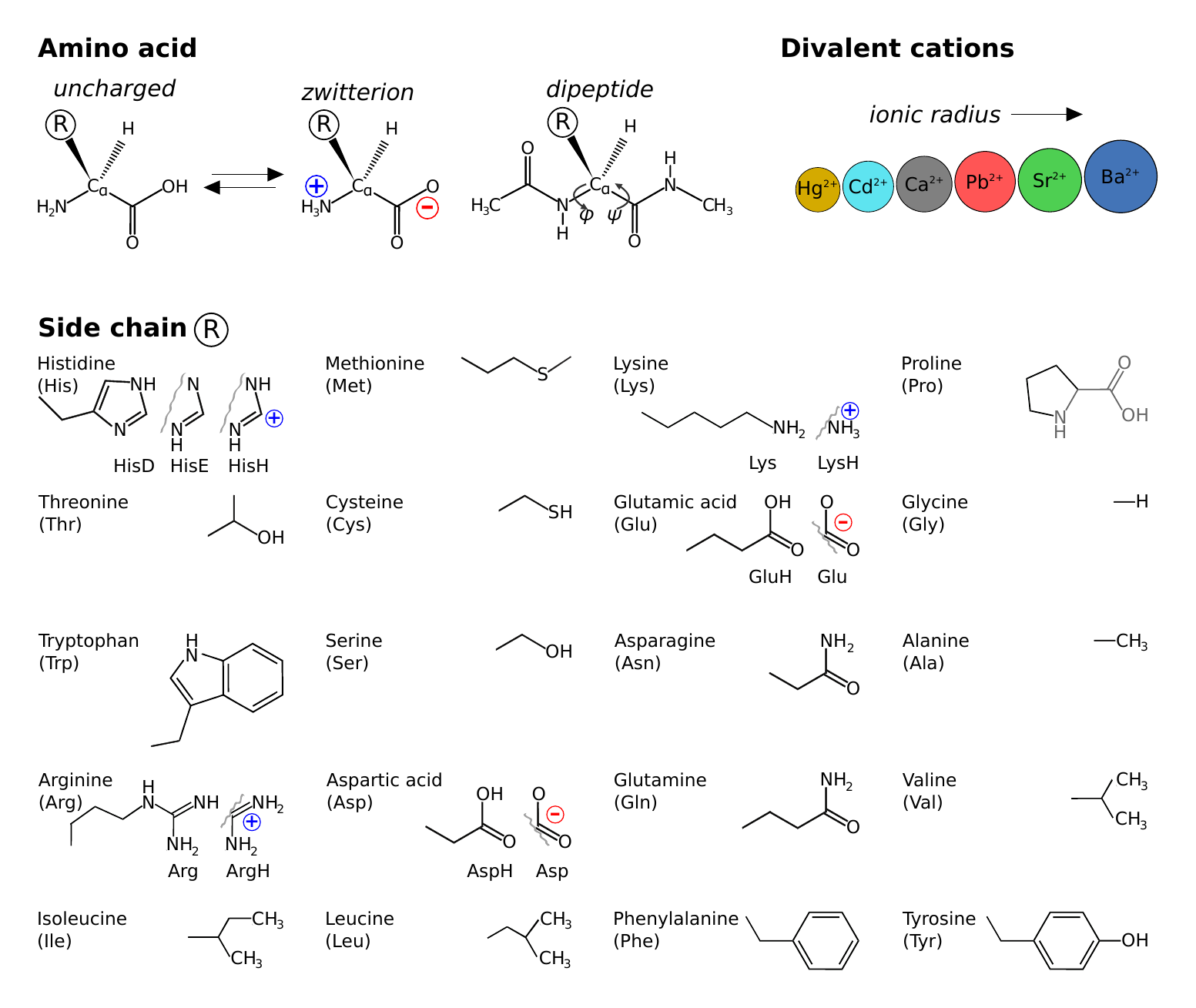}
\caption{{\bf Molecular systems covered in this study.} 
Top left and center: Schematic depiction of the backbone conformations of uncharged, zwitterionic, and dipeptide forms of the aminoacids considered in this work. Side chains are indicated by the letter \textsl{\textbf{R}}. Top right: Divalent ions considered for complexation with the 20 proteinogenic amino acids. Lower five rows: Side chains, including different protonation states where applicable, of the 20 proteinogenic amino acids considered in this work.
} 
\label{fig:AA_scheme} 
\end{figure}

Proteins are the machinery of life.
We here present a first-principles study of the conformational preferences of their basic building blocks -- specifically, as summarized in Figure~\ref{fig:AA_scheme}: 20 proteinogenic amino acids and dipeptides, with different possible protonation states, and the conformational space changes resulting from attaching six divalent cations, i.e., Ca$^{2+}$, Ba$^{2+}$, Sr$^{2+}$, Cd$^{2+}$, Pb$^{2+}$, and Hg$^{2+}$. 
In past studies, a wide range of different approximate electronic structure methods has been applied to some of these proteinogenic amino acids -- see, for example, references \cite{jcc30_2105, mp106_2289, pbmb71_243, msr31_391, psfb48_107,jcc29_407,jms346_141,jacs114_9568,pccp13_18561,pccp11_3921,jpca102_5111,jcc28_1817,jmst671_77,jmst332_251,jctc6_3066,jacs119_5908,jpca115_9658,jcp127_154314,jcp137_75102,jpca112_3319,ijms283_56,jcc18_1609,jpca114_5919,jpca116_3247,jpca115_2900,jpca114_7583,jpca109_2660,cpl453_1,ijqc112_1526,jcp118_1253,jcp122_134313,jmst953_28,jmst631_277,jmst719_153,jmst666_273,jmst540_271,sapa73_865,pccp10_1248,pccp15_6097,pccp12_4899,pccp9_4698,prl91_203003,pnas104_20183,mp107_761,jpcb116_12441,ctc976_42,jpoc24_553,jpoc20_1099,jacs115_2923,jctc9_1533,njc29_1540,baldauf2012ab,ijqc65_1033,cej9_1008,jpc100_11589,cpb20_033102,Yuan2014,Karton2014,Kesharwani2015}. 
These studies have deepened our understanding of the conformational basics of individual building blocks, but a systematic comparison of properties of the different building blocks is complicated when relying on data from different sources.
On the one hand this is due to the molecular models that may differ in protonation states and backbone capping.
On the other, the simulations can differ in several ways: 
\begin{itemize}
\item Different sampling strategies or methods to generate conformers may have been used. Search-dependent settings, like energy cut-offs, can also have a significant impact on the results.  
\item The levels of theory that have been applied range from semi-empirical to Hartree-Fock (HF) to density-functional theory (DFT) up to coupled-cluster calculations \cite{jcc30_2105, mp106_2289, pbmb71_243, msr31_391, psfb48_107,jcc29_407,jms346_141,jacs114_9568,pccp13_18561,pccp11_3921,jpca102_5111,jcc28_1817,jmst671_77,jmst332_251,jctc6_3066,jacs119_5908,jpca115_9658,jcp127_154314,jcp137_75102,jpca112_3319,ijms283_56,jcc18_1609,jpca114_5919,jpca116_3247,jpca115_2900,jpca114_7583,jpca109_2660,cpl453_1,ijqc112_1526,jcp118_1253,jcp122_134313,jmst953_28,jmst631_277,jmst719_153,jmst666_273,jmst540_271,sapa73_865,pccp10_1248,pccp15_6097,pccp12_4899,pccp9_4698,prl91_203003,pnas104_20183,mp107_761,jpcb116_12441,ctc976_42,jpoc24_553,jpoc20_1099,jacs115_2923,jctc9_1533,njc29_1540,baldauf2012ab,ijqc65_1033,cej9_1008,jpc100_11589,cpb20_033102,Yuan2014,Karton2014,Kesharwani2015}.
\item Numerical settings, e.g., basis sets, can differ substantially and might lead to different results.
\end{itemize}

A further point that limits a quantitative comparison is the accessibility of the data from different studies.
Energies, for example, often have to be extracted from table footnotes and/or the structural data is not always accessible in the Supporting Information of the respective articles, sometimes even only accessible as figures in the manuscript.
The data set presented here overcomes such limitations by covering a comprehensive segment of chemical space exhaustively, using a large scale computational effort.
This study treats 20 proteinogenic amino acids, their dipeptides and their interactions with the divalent cations Ca$^{2+}$, Ba$^{2+}$, Sr$^{2+}$, Cd$^{2+}$, Pb$^{2+}$, and Hg$^{2+}$ (see Figure~\ref{fig:AA_scheme} for an overview) on the same theoretical footing.
The importance of peptide cation interactions may be highlighted by the fact that about 40\% of all proteins bind cations\cite{cr96_2239,cob3_378,jib102_1901}.
Especially Ca$^{2+}$ is important in a multitude of functions, ranging, for example, from blood clotting\cite{zhou2011novel} to cell signaling to bone growth\cite{ebj39_825}. 
Such calcium mediated functions can be disturbed by the presence of alternative divalent heavy metal cations like Pb$^{2+}$, Cd$^{2+}$, and Hg$^{2+}$\cite{jt132671,jib102_1901,bbrc372_341}.

The conformations and total energies of each molecular system are calculated from first principles in the framework of density-functional theory (DFT) \cite{pr136_b864,pr140_a1133} using the PBE generalized-gradient exchange-correlation functional\cite{prl77_3865}.
Energies are corrected for van der Waals interactions using the Tkatchenko-Scheffler formalism \cite{prl102_73005}.
In this formalism, pairwise $C_6 [n]/r^6$ terms are computed and summed up for all pairs of atoms. $r$ is the interatomic distance, a cut-off for short interatomic distances is applied, and $C_6 [n]$ coefficients are obtained from the self-consistent electron density.
The combined approach is referred to as ``PBE+vdW'' throughout this work.
This level of theory is robust for potential-energy surface (PES) sampling of peptide systems \cite{jpcl1_3465,prl106_118102,cej19_11224,doi:10.1021/jp3098268,doi:10.1021/jp402087e,doi:10.1021/jp412055r,C4CP05216A,C4CP05541A}.
The curated data is provided as basis for comparative studies across chemical space to reveal conformational trends and energetic preferences.
It can, for example, further be used for force-field development, theoretical studies at higher levels of theory, and as a starting point for theoretical calculations of spectra for biophysical applications.

\section*{Methods }
\subsection*{Molecular models}
This study covers a total of 280 molecular systems (summarized in Figure~\ref{fig:AA_scheme}). 
The number is the product of these chemical degrees of freedom that were considered in our study:
\begin{description}
\item[20] proteinogenic amino acids. In case of (de)protonatable side chains, all protomers (different protonations states) were considered as well.
\item[2] different backbone types, either free termini (considered in uncharged or zwitterionic form) or capped (N-terminally acetylated or C-terminally amino-methylated).
\item[7] reflecting that the respective amino acid or dipeptide was considered either in isolation or with one of six different cation additions: Ca$^{2+}$, Ba$^{2+}$, Sr$^{2+}$, Cd$^{2+}$, Pb$^{2+}$, or Hg$^{2+}$.
\end{description}

\subsection*{Conformational search and energy functions}
\begin{figure}
\begin{center} 
\includegraphics[width=0.5\textwidth]{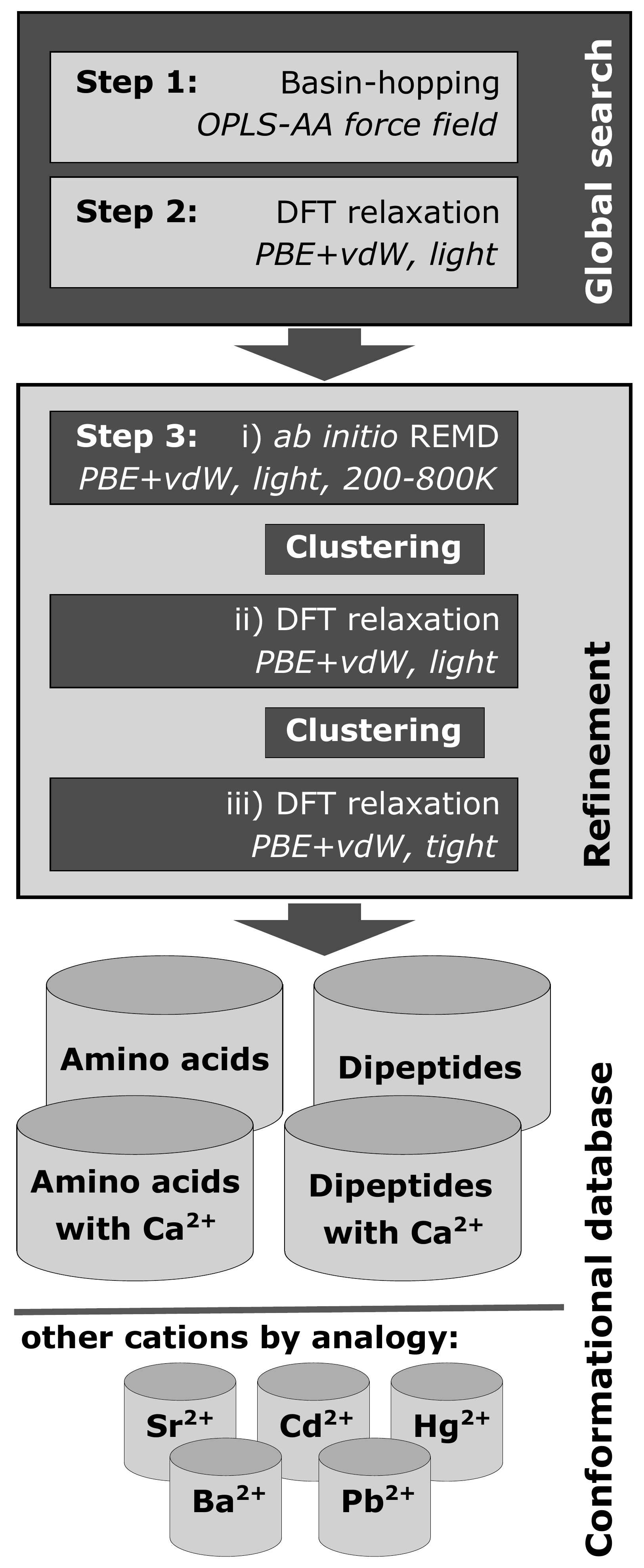}
\end{center}
\caption{{\bf Schematic representation of the workflow} employed to locate stationary points on the potential-energy surfaces of the respective molecular systems.} 
\label{fig:workflow} 
\end{figure}

For the initial scan of the PES, the empirical force field OPLS-AA \cite{jacs118_11225} was employed, followed by DFT-PBE+vdW relaxations of the energy minima identified in the force field. 
The identified set of structures was then subjected to a further first-principles refinement step, \textit{ab initio} replica-exchange molecular dynamics (REMD). 
An overview of the procedure is given in Figure~\ref{fig:workflow} and the steps are described in more detail below.

Force-field based (OPLS-AA) \cite{jacs118_11225} \textbf{global conformational searches (Step~1)} were performed for all dipeptides and amino acids (i) without a coordinating cation and (ii) with Ca$^{2+}$.
These searches employed a basin hopping search strategy\cite{jpca101_5111, science285_1368} as implemented in the tool ``scan'', distributed with the \textsc{Tinker} molecular simulation package \cite{jcc87_1016,jpcb107_5933}. 
We here use an in-house parallelized version of the \textsc{Tinker} scan utility that was first used in reference \cite{doi:10.1021/jp3098268}.
In this search strategy, input structures for relaxations are generated by projecting along normal modes starting from a local minimum.
The number of search directions from a local minimum was set to 20.
Conformers were accepted within a relative energy window of 100\,kcal/mol and if they differ in energy from already found minima by at least 10$^{-4}$\,kcal/mol.
The search terminates when the relaxations of input structures do not result in new minima.

After that, \textbf{PBE+vdW relaxations (Step~2)} were performed with the program FHI-aims \cite{cpc180_2175,Havu20098367,1367-2630-14-5-053020}.
FHI-aims employs numeric atom-centered orbital basis sets as described in reference \cite{cpc180_2175} to discretize the Kohn-Sham orbitals.
Different levels of computational defaults are available, distinguished by choice of the basis set, integration grids, and the order of the multipole expansion of the electrostatic (Hartree) potential of the electron density.
For the chemical elements relevant to this work, “light” settings include the so-called \textit{tier1} basis sets and were used for initial relaxations.
“Tight” settings include the larger \textit{tier2} basis sets and ensure converged conformational energy differences at a level of few meV \cite{cpc180_2175}.
Unless noted otherwise, all energies discussed here are results of PBE+vdW calculations with a \textit{tier2} basis and “tight” settings.
Relativistic effects were taken into account by the so-called atomic zero-order regular approximation (atomic ZORA)\cite{cpl328_107,jcp109_392} as described in reference \cite{cpc180_2175}.
Previous comparisons to high-level quantum chemistry benchmark calculations at the coupled-cluster level, CCSD(T), demonstrated the reliability of this approach for polyalanine systems \cite{prl106_118102,doi:10.1021/jp412055r}, alanine, phenylalanine, and glycine containing tripeptides \cite{doi:10.1021/jp412055r}, and alanine dipeptides with Li$^+$ \cite{cej19_11224}. Further benchmarks at the MP2 level of theory are reported below in the section Technical Validation.

\begin{figure}
\begin{center} 
\includegraphics[width=1.2\textwidth]{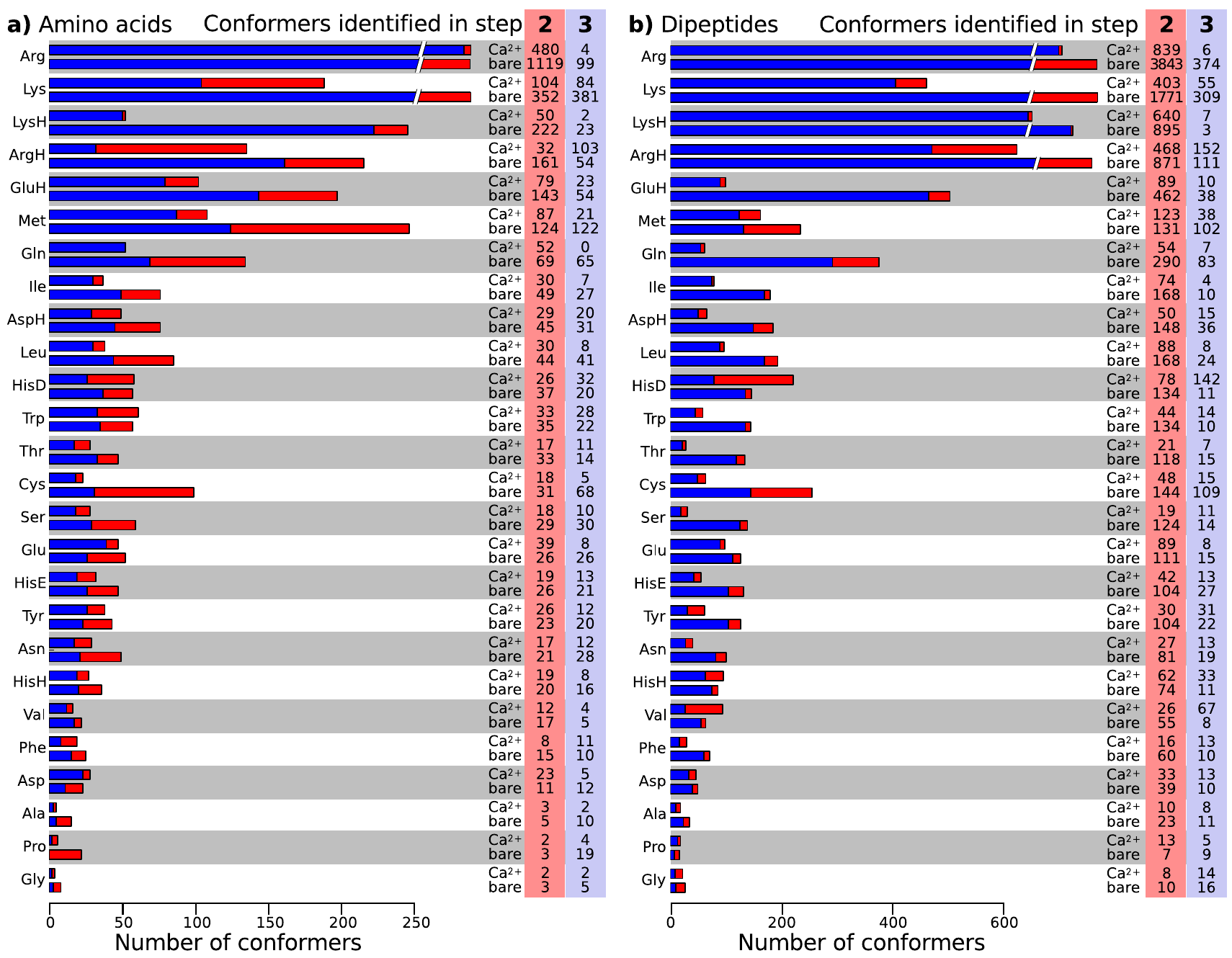}
\end{center}
\caption{{\bf Numbers of stationary points} of the PBE+vdW potential-energy surface (PES) at the ``tight''/tier-2 level of accuracy that were found for the different \textbf{a)} uncapped amino acids or \textbf{b)} dipeptides in isolation (``bare'') or with a Ca$^{2+}$ cation.
Blue segments of the bars and blue shaded numbers give the number of stationary points (``conformers'') located in Step 2 of the search procedure detailed in Figure~\ref{fig:workflow}. Red bar segments and red shading highlight the number of conformers that were additionally found during Step 3 of the search. The total number of conformers found for each system is the sum of the numbers found in steps two and steps three.} 
\label{fig:HowMany} 
\end{figure}

The \textbf{refinement (Step~3)} by \textit{ab initio} REMD\cite{prl57_2607,cpl314_141} is intended to alleviate the potential effects of conformational energy landscape differences between the force field and the DFT method.
In REMD, multiple molecular dynamics trajectories of the same system are independently initialized and run in a range of different temperatures.
Based on a Metropolis criterion, configurations are swapped between trajectories of neighboring temperatures. 
Thus, the simulations can overcome barriers and provide an enhanced conformational sampling in comparison to classical molecular dynamics (MD)\cite{pccp7_3910,cpl314_141}. 
The simulations were carried out employing a script-based REMD scheme that is provided with FHI-aims and that was first used in reference \cite{C1FD00027F}.
Computations were performed at the PBE+vdW level with “light” computational settings. 
The run time for each REMD simulation was 20\,ps with an integration time step of 1\,fs. 
The frequent exchange attempts (every 0.04 or 0.1\,ps) ensure efficient sampling of the potential-energy surface as shown by Sindhikara \emph{et al.}\cite{jcp128_24103}. 
The velocity-rescaling approach by Bussi \emph{et al.}\cite{jcp126_14101} was used to sample the canonical distribution. 
Starting geometries for the replicas were taken from the lowest energy conformers resulting from the PBE+vdW relaxations in Step 2.  
REMD parameters for the individual systems, i.e. the number of replicas, acceptance rates for exchanges between replicas, the frequency for exchange attempts, and the temperature range, are summarized in table S1 of the Supporting Material.
Conformations were extracted from the REMD trajectories every 10th step, i.e. every 10\,fs of simulation time. 
In order to generate a set of representative conformers, these structures were clustered using a $k$-means clustering algorithm\cite{as28_100} with a cluster radius of 0.3\,\AA{} as provided by the MMSTB package\cite{jmgm22_377}. 
The resulting arithmetic-mean structures from each cluster were then relaxed using PBE+vdW with “light” computational settings. 
The obtained conformers were again clustered and cluster representatives were relaxed with PBE+vdW (“tight” computational settings) to obtain the final conformation hierarchies. 
The refinement step by REMD is essential, as shown in Figure~\ref{fig:HowMany}, which separately identifies the number of distinct conformers found in Step~2 and, subsequently, the number of additional conformers found in Step~3.

After step 2, a total of 17,381 stationary points was found for the amino acids and dipeptides in isolation and in complex with Ca$^{2+}$. The refinement procedure in Step 3 increases this number to a total of 21,259 structures. Initial structures for the Ba$^{2+}$, Cd$^{2+}$, Hg$^{2+}$, Pb$^{2+}$ and Sr$^{2+}$ binding amino acid and dipeptide systems were then obtained by replacing the Ca$^{2+}$ cation in the amino acid and dipeptide structures binding a Ca$^{2+}$ cation. 
These structures were subsequently relaxed with PBE+vdW employing “tight” computational settings and a tier-2 basis set.
This procedure results in 24,633 further conformers with bound cations. Altogether, we thus provide information on 45,892 stationary points of the PBE+vdW PES for all systems studied in this work.

The numbers of conformers identified in the searches are also given in Table S2 of the Supporting Material. Tables S3 and S4 provide detailed accounts of how many structures were found for which amino acid/dipeptide in isolation or with attached cations. 

\section*{Data Records }
The curated data, consisting of the Cartesian coordinates of 45,892 stationary point geometries of the PBE+vdW PES (the main outcome of our work) and their potential energies computed at the ``tight''/tier-2 level of accuracy in the FHI-aims code, is provided as plain text files sorted in directories (see Figure~\ref{fig:folders}). The PBE+vdW total energies are included since they are an integral part of the construction of our geometry data sets. Importantly, the stationary point geometries could be used as starting points to refine the total energy accuracy by higher-level methods, e.g., those discussed in ``Technical Validation'' below.
The folder structure is hierarchic and straightforward.
The naming scheme is explained in the following:

Description of the file types:
\begin{description}
\item[conformer.(...).xyz] coordinates in standard xyz format in \AA{}, readable by a wide range of molecule viewers, e.g. VMD, Jmol, etc.
\item[conformer.(...).fhiaims] coordinate file in FHI-aims geometry input format: for each atom of the particular system, the Cartesian coordinates are given in \AA{} (\texttt{atom [x] [y] [z] [element]}). The electronic total energy (in eV) at the PBE+vdW level is given there as a comment.
\item[control.in] FHI-aims input file with technical parameters for the calculations. Please note that these files also include the exact specifications of the “tight” numerical settings for all included elements.
\item[hierarchy\_PBE+vdW\_tier-2.dat] in each final subfolder, contains three columns: number of the conformer, total energy (in eV, PBE+vdW, tier-2 basis set, “tight” numerical settings, computed with FHI-aims version 031011), and relative energy (in eV, relative to the respective global minimum).
\end{description}

The curated data is publicly available from several sources:
\begin{enumerate}
\item A website dedicated to this data set has been set up\footnote{http://aminoaciddb.rz-berlin.mpg.de} and allows users to browse and download the data and to visualize molecular structures online.
\item From the NOMAD repository\footnote{http://nomad-repository.eu} the data is available via the DOI 10.17172/NOMAD/20150526220502\footnote{http://dx.doi.org/10.17172/NOMAD/20150526220502} [Data citation 1].
\item In addition, the data has been uploaded to DRYAD\footnote{https://datadryad.org} and has been assigned the DOI 10.5061/dryad.vd177\footnote{http://dx.doi.org/10.5061/dryad.vd177} [Data citation 2].
\end{enumerate}

\section*{Technical Validation }
\begin{figure}
\includegraphics[width=1\textwidth]{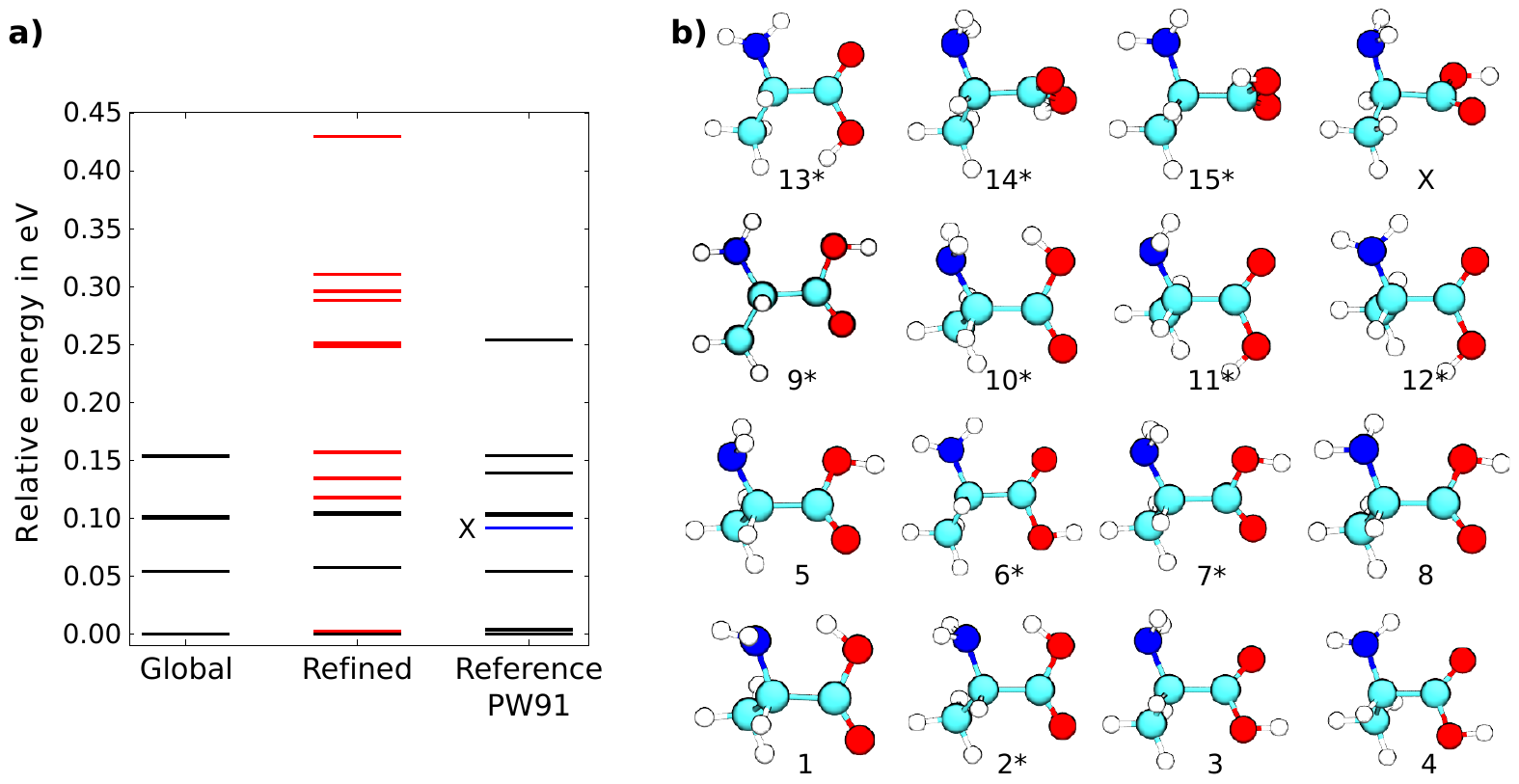} 
\caption{{\bf Comparison of search strategies. }
(\textbf{a)}) The conformational energy hierarchies for alanine after the global search and the local refinement together with the reference hierarchy at the DFT-PW91 level that was published by Maul \textit{et al}.\cite{jcc28_1817}. Conformers indicated by black lines were found in the global search, the conformers in red were located only after the local refinement step. The blue line in the reference conformational hierarchy represents a minimum not found in our search and not present at the PBE+vdW level.
(\textbf{b)}) Conformations of the alanine molecule. Conformers marked with an asterisk (*) were found in the local refinement step of our search strategy. Atoms are color-coded as follows: Cyan (C), blue (N), red (O), white (H).
The conformer labeled with X was found by Maul \emph{et al.} in PW91 calculations\cite{jcc28_1817} but is unstable at the PBE+vdW level.
}
\label{ala_example}
\end{figure}

The conformational coverage for the amino acid alanine is validated by comparing to a recent study by Maul \textit{et al.}\cite{jcc28_1817} .
In that reference, 10 low energy conformers of alanine were reported, spanning an energy range of approximately 0.26\,eV between the reported lowest and highest energy conformers.
The level of theory used by Maul \textit{et al.} was DFT in the generalized gradient approximation by means of the Perdew-Wang 1991 functional\cite{Perdew91}.
In our case, the force field based search step with subsequent PBE+vdW relaxations yields 5 conformers.
The following \textit{ab initio} REMD simulations increase the number of conformers to 15 within an energy range of 0.43\,eV.
The respective conformational energy hierarchies after global search and after REMD-refinement are shown in Figure~\ref{ala_example}A. 
The results of our search (with the refinement step) are in good agreement with the data from reference \cite{jcc28_1817} that is also shown in Figure~\ref{ala_example}A.
Structures are shown in Figure~\ref{ala_example}B.
Nine of the ten conformers identified by Maul \textit{et al.} can be confirmed. 
The single conformer that is missing (highlighted by an X in Figure~\ref{ala_example}A) is not a stationary point of the PBE+vdW potential energy surface. 
Conformers 14 and 15 are classified as saddle points by analysis of the vibrational modes.

\begin{figure}
\begin{center}
\includegraphics[width=1.0\textwidth]{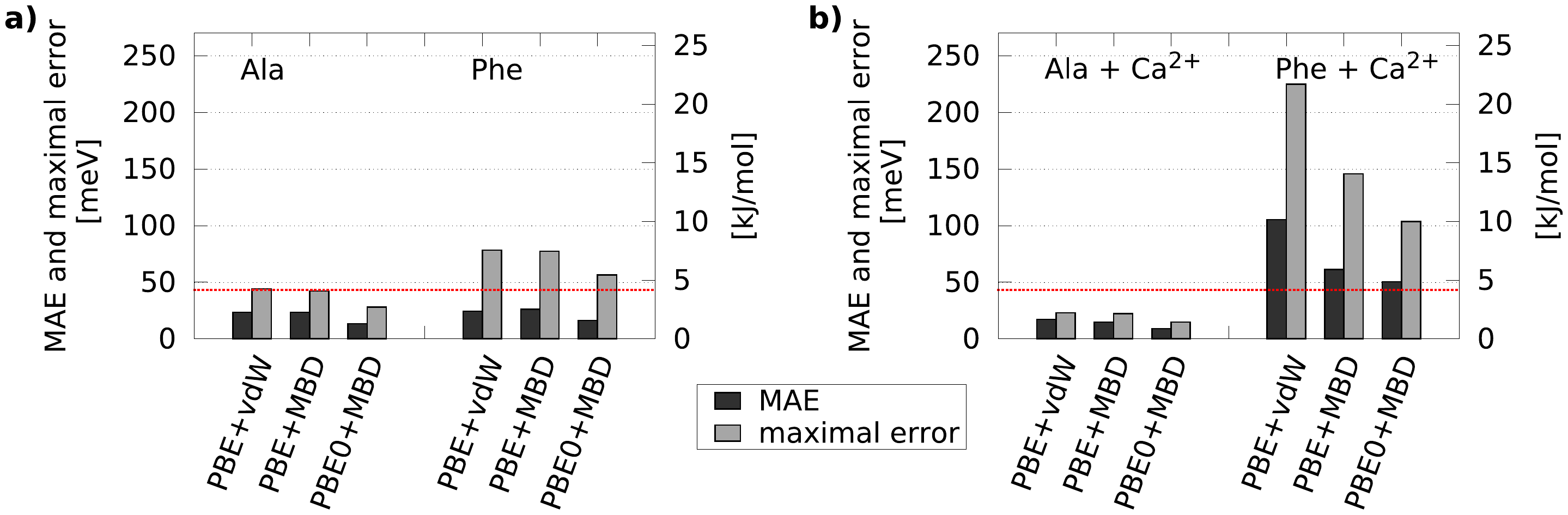} 
\end{center}
\caption{{\bf Comparison of different DFAs to MP2 energies.}
Mean absolute error (MAE) and maximal error (in meV) between different relative energies at the DFA (PBE+vdW, PBE+MBD*, and PBE0+MBD*) and MP2 level of theory, using structures of obtained minima on the PBE+vdW level from the database for the systems of Ala and Phe with neutral end caps, both in isolation and in complex with a Ca$^{2+}$ cation. Computational details are given in the text. Exact numbers are summarized in Table~\ref{tbl:rmsd-dfa-vs-mp2}. 
}
\label{MP2bench}
\end{figure}

In order to further quantify the reliability of the DFT-PBE+vdW level of theory for peptides, beyond earlier benchmark work\cite{prl106_118102,doi:10.1021/jp412055r,cej19_11224} and especially with divalent cations, benchmark calculations were performed at the level of M{\o}ller-Plesset second-order perturbation theory (MP2) \cite{MP2-1,MP2-2} using the electronic structure program package ORCA \cite{ORCA}.
Single-point energy calculations were performed for all fixed stationary-point DFT-PBE+vdW geometries in our data base for the amino acids alanine (Ala) and phenylalanine (Phe) with neutral N and C termini in isolation as well as in complex with a Ca$^{2+}$ cation. Phe was selected to represent a ``difficult'' example, i.e., the interaction of the cation with a larger aromatic side chain. 
The MP2 calculations did not include any frozen-core treatment (including semicore states is essential for Ca$^{2+}$) and were performed using Dunning's correlation-consistent polarized core-valence basis sets (cc-pCVnZ), with n=T/Q/5 denoting the triple-zeta, quadruple-zeta, and quintuple-zeta basis sets respectively \cite{cc-pCVnZ}. 
The calculated SCF (Hartree-Fock) and MP2 correlation energies were then individually extrapolated to the complete basis set (CBS) limit as follows: For SCF energies, we used the extrapolation strategy proposed by Karton and Martin \cite{SCFextrapolation}:
\begin{equation} 
E^{n}_{SCF}=E^{CBS}_{SCF}+A e^{-\alpha\sqrt{n}}.
\label{CBSSCF}
\end{equation}\\
$A$, $\alpha$, and the CBS-extrapolated energy $E^{CBS}_{SCF}$ are parameters determined from a least-squares fitting algorithm applied individually for each conformer. 
For the MP2 correlation energies, an extrapolation scheme proposed by Truhlar \cite{MP2extrapolation} was applied:
\begin{equation} 
E^{n}_{corr}=E^{CBS}_{corr}+B n^{-\beta}.
\label{CBSMP2}
\end{equation}\\
Again, $B$, $\beta$, and the CBS-extrapolated energy $E^{CBS}_{corr}$ are parameters determined from a least-squares fitting algorithm as before. A detailed account of all numbers is given in the Supporting Material (Table S5).
Mean absolute errors between the density-functional approximation (DFA) relative energies and the basis-set extrapolated MP2 relative energies were calculated as follows:
\begin{equation} 
MAE = \frac{1}{N} \sum_{i=1}^{N} |\Delta E_i^{DFA} - \Delta E_i^{MP2}+c|,
\label{EquMAE}
\end{equation}\\
where the index $i$ runs over all $N$ conformations of a given data set. 
$\Delta E_i$ in principle denotes the energy difference between conformer $i$ and the lowest-energy conformer of the set. 
The adjustable parameter $c$ is used to shift the MP2 and DFA conformational hierarchies versus one another to obtain the lowest possible MAE, rendering the reported MAE value independent of the choice of any reference structure. 
Figure~\ref{MP2bench}A shows the corresponding obtained mean absolute errors (MAE) and maximal errors ($max_{i}|\Delta E_i^{DFA} - \Delta E_i^{MP2}+c|$) of different DFA calculations -- performed with the FHI-aims code -- with respect to benchmarks on the MP2 level obtained as described above. 
Within FHI-aims, the accuracy of integration grids and of the electrostatic potential was also verified by comparing ``tight'' and ``really\_tight'' numerical settings, giving virtually identical results.
The DFA level of theory of PBE+vdW shows a MAE well within \textit{chemical accuracy} of $\sim1\,\mathrm{kcal/mol}\approx43\,\mathrm{meV}$ for both structural sets of Ala and Phe; for Phe, the maximal error is $\sim2\,\mathrm{kcal/mol}$.
We next applied a different long-range dispersion treatment, a recent many-body dispersion model based on interacting quantum harmonic oscillators denoted as MBD,\cite{MBD} showing no significant improvement for the isolated amino acids. In line with Ref. \cite{doi:10.1021/jp412055r},
applying the more expensive PBE0 \cite{PBE0Adamo} hybrid exchange correlation functional reduces the maximum deviation for Phe to $\sim57\,\mathrm{meV}$, i.e., 1.3\,kcal/mol.
For Ala and Phe with neutral end caps in complex with a Ca$^{2+}$ cation, Figure~\ref{MP2bench}B compares the same set of DFAs to MP2 benchmark energy hierarchies. However, obtaining basis-set converged total energies of the same accuracy as for the isolated peptides by straightforward CBS extrapolation proved remarkably more difficult when Ca$^{2+}$ was involved. The reason is traced to the significant and slow-converging correlation contribution of the Ca$^{2+}$ semicore electrons, which leads to large and conformation dependent basis set superposition errors (BSSE). This problem was verified for MP2 calculations in the FHI-aims and ORCA codes, with several different basis set prescriptions \cite{Zhang2013}, and for CCSD(T) calculations. Standard DFAs, if sufficiently accurate, have a significant advantage in this respect since they are not subject to comparable numerical convergence problems. To yet arrive at reliable CBS-extrapolated MP2 conformational energy differences, we thus subjected the SCF and correlation energies of each Ca$^{2+}$ coordinated conformation to a counterpoise correction\cite{counterpoise1,counterpoise2} to minimize the effect of BSSE on the Ca$^{2+}$ correlation energy contribution, prior to performing CBS extrapolation as described above.
For the example of Ala+Ca$^{2+}$ and assuming rigid conformers, the BSSE  is estimated as:

\begin{align}
\begin{split}
E_{BSSE}         =&E_{BSSE}(Ala)+E_{BSSE}(Ca^{2+})\text{ , with}\\
&E_{BSSE}(Ala)    =E^{Ala+Ca^{2+}}(Ala)-E^{Ala}(Ala)\text{ , and}\\
&E_{BSSE}(Ca^{2+})=E^{Ala+Ca^{2+}}(Ca^{2+})-E^{Ca^{2+}}(Ca^{2+}) .
\end{split}
\label{BSSE}
\end{align}

$E^{Ala+Ca^{2+}}(Ala)$ represents the energy of Ala evaluated in the union of the basis sets on Ala and Ca$^{2+}$, $E^{Ala}(Ala)$ represents the energy of Ala evaluated in the basis set on Ala, \textit{etc.} The individual BSSE errors are then subtracted from the SCF and correlation energy respectively. Phe+Ca$^{2+}$ is treated equivalently. Complete numerical details are given in the Supplementary Material (Table S6).
Following this procedure, the MAE and maximal error values of various DFAs compared to MP2 are well within 1\,kcal/mol for Ala+Ca$^{2+}$. 
The PBE+vdW MAE for Phe+Ca$^{2+}$ amounts to just above $\sim2\,\mathrm{kcal/mol}$. The contributions from both the many-body dispersion and the hybrid PBE0 functional improve the MAE to just above $1\,\mathrm{kcal/mol}$ at to PBE0+MBD* level of theory. The maximum errors in the energy hierarchies between individual conformers are correspondingly larger. Overall, this assessment shows that our data base of conformer geometries constitutes, e.g., an excellent starting point for more exhaustive future benchmark work of new electronic structure methods for cation-peptide systems. For example, it would be very interesting to explore how F12 approaches, which address the correlation energy convergence problem explicitly, fare for a broad range of different Ca$^{2+}$ containing conformations of our peptides. 

\begin{figure}
\begin{center}
\includegraphics[width=0.8\textwidth]{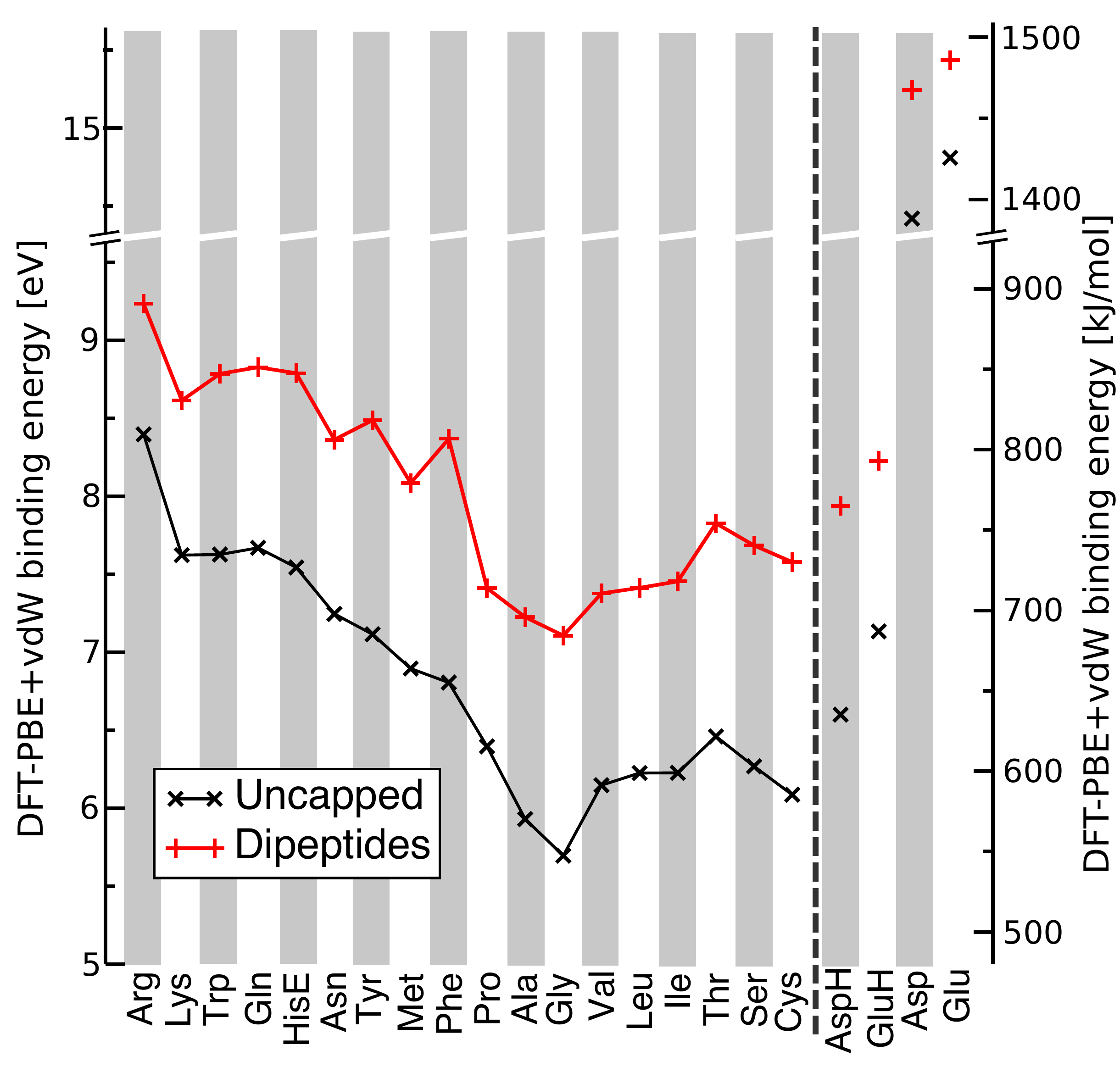}
\end{center}
\caption{{\bf Comparison of the gas-phase binding energies of Ca$^{2+}$ to different amino acids} calculated in this work ($y$ axis) to the experimentally inferred hierarchy of gas-phase binding energies of Ca$^{2+}$ to different amino acids by Ho \emph{et al.}\cite{rcms21_1097} The amino acids are ordered along the $x$ axis from the highest to lowest experimental Ca$^{2+}$ binding energy. Protonated and deprotonated Asp and Glu are not included among the experimental data and are here shown as predictions. $E_\mathrm{binding}$ is high for deprotonated Asp and Glu since these forms of the amino acid would carry a negative charge.} 
\label{be_exp} 
\end{figure}

As a final validation, we compare the correlation of calculated gas-phase amino acid-Ca$^{2+}$ binding energies to the binding energy hierarchy found experimentally in a study by Ho \emph{et al.}\cite{rcms21_1097}. We calculate binding energy at the PES level as
\begin{equation} 
E_{binding} =  E_{amino\,\,acid} + E_{cation} - E_{complex} .
\label{Ebind}
\end{equation}
Energies $E$ denote the PBE+vdW Born-Oppenheimer potential energies, including $E_{amino\,\,acid}$ of the lowest-energy conformers of the isolated amino acid and $E_{complex}$ of the same amino acid in complex with a Ca$^{2+}$ ion. Experimentally \cite{rcms21_1097}, the gas-phase Ca$^{2+}$ affinities of 18 proteinogenic amino acids were determined by fragmenting Ca$^{2+}$ complexes with a combinatoric library of tripeptides at $T\approx$330~K, recording the mass spectrometric peak intensities of different fragmentation products. Quantitative average relative binding energies of $Ca^{2+}$ to different amino acids were thus inferred and can be compared to our findings, albeit with several important experiment-theory differences: (i) Entropy effects \cite{Liwo15022005,cej19_11224,doi:10.1021/jp402087e} should affect the specific complexes probed experimentally but cannot be included into the calculated numbers in the exact same way, (ii) structural differences (e.g., protonation, dimerized amino acids) between the fragments recorded in experiment and the amino acids covered here, (iii) experimental $Ca^{2+}$ affinities are not given for Asp and Glu because their gas-phase acidities, needed for data conversion, are not known. 
Figure~\ref{be_exp} compares the experimentally and theoretically inferred $Ca^{2+}$ binding affinities qualitatively. 
The $x$-axis reflects the experimental binding affinity energy hierarchy, arranging amino acids from left to right in order of decreasing binding affinity. 
The $y$ axis shows calculated binding energies according to Eq.~\ref{Ebind}. 
Perfect correlation of the experimental and calculated hierarchies would imply a strictly monotonic decrease of calculated $E_{binding}$ values from left to right. 
This monotonic trend is not obeyed exactly; however, in view of the significant differences (i) and (ii) above, the qualitative agreement is quite striking. 
Normalized correlation coefficients between the experimental (1) and calculated (2) binding affinity data were calculated following the formula:
\begin{equation} 
r_{12} = s_{12}/(s_1 s_2), 
\label{corrcoeff}
\end{equation}
with $s_{12}$ being the covariance of data sets and $s_i$ being the standard deviations of data sets $i$=1,2.
The result, correlation coefficients of $r_{12}$=0.979 or 0.909 for uncapped amino acids or dipeptides, respectively, also point to an overall remarkably good agreement.
Finally, Figure~\ref{be_exp} also gives predicted $E_{binding}$ values for protonated (overall system charge +2) and deprotonated (overall system charge +1) Asp and Glu, reflecting the significant electrostatic attraction between cations and negatively charged (deprotonated) Asp and Glu side chains. 
The binding energy data sets are included as Supplementary Table S5.

\section*{Usage Notes}
The present data contains stationary-point geometries (mainly minima, but also saddle points since no routine normal-mode analysis was performed) on the potential energy surface of the 20 proteinogenic amino acids and dipeptides, either isolated or in complex with a divalent cation (Ca$^{2+}$, Ba$^{2+}$, Sr$^{2+}$, Cd$^{2+}$, Pb$^{2+}$, Hg$^{2+}$).
The users of this dataset may find openbabel\cite{joc3_33}(www.openbabel.org) to be a useful tool to convert FHI-aims and xyz files to other common file formats in chemistry. 

\section*{Author Contributions}
MR performed the calculations to assemble all conformers. 
MR and CB curated the data.
Validation calculations by DFAs and correlated methods other than PBE+vdW were carried out by MS.
MR, CB, VB designed the study and wrote the data descriptor.

\section*{Acknowledgements }
The authors are grateful to Matthias Scheffler (Fritz Haber Institute Berlin) for support of this work and stimulating discussions.
Luca Ghiringhelli is gratefully acknowledged for his work on the script-based parallel-tempering scheme that is provided with FHI-aims and that was used in the present work. The authors thank Robert Maul and Karsten Hannewald for making available the original alanine geometries derived in their 2007 study for comparison with the present results. The authors further thank Mariana Rossi, Franziska Schubert, and Sucismita Chutia for sharing their extensive experience with all search methods employed in this work.

\section*{Competing financial interests }
The authors declare no competing financial interests.

\clearpage
\section*{Figures and Legends}

\begin{figure}
\dirtree{%
.1 AA-Dataset.
.2 Ala.
.2 Arg.
.2 ArgH.
.2 Asn.
.2 Asp.
.2 AspH.
.2 Cys.
.3 uncapped.
.3 dipeptide.
.4 bare.
.4 Ba.
.4 Ca.
.5 \textit{hierarchy\_PBE+vdW\_tier-2.dat}.
.5 \textit{control.in}.
.5 \textit{conformer.0001.fhiaims}.
.5 \textit{conformer.({...}).fhiaims}.
.5 \textit{conformer.0001.xyz}.
.5 \textit{conformer.({...}).xyz}.
.4 Cd.
.4 Hg.
.4 Pb.
.4 Sr.
.2 Gln.
.2 Glu.
.2 GluH.
.2 Gly.
.2 HisD.
.2 HisE.
.2 HisH.
.2 Ile.
.2 Met.
.2 Leu.
.2 Lys.
.2 LysH.
.2 Phe.
.2 Pro.
.2 Ser.
.2 Thr.
.2 Trp.
.2 Tyr.
.2 Val.
}
\caption{{\bf Schematic representation of folder organization of the data.} Each folder, as exemplified for the Ca$^{2+}$-coordinated cysteine dipeptide, contains coordinate files in two formats (standard XYZ and FHI-aims input), the computational settings file for FHI-aims (control.in), and the energy hierarchies (PBE+vdW, “tight”/tier-2 level) per system.} 
\label{fig:folders} 
\end{figure}

\clearpage

\section*{Tables}

\begin{table}
\caption{Mean absolute error (MAE) and maximal error (in meV; in parentheses: in kcal/mol) between different relative energies at the DFA (PBE+vdW, PBE+MBD*, and PBE0+MBD*) and MP2 level of theory, using structures of obtained minima on the PBE+vdW level from the database for the systems of Ala and Phe with neutral end caps, both in isolation and in complex with a Ca$^{2+}$ cation. Computational details are given in the text.}
\centering
\begin{tabular}{|c|c|c|c|}
\hline
\multicolumn{2}{|c|}{\textbf{System}} & \textbf{MAE} [meV] & \textbf{Maximal error} [meV] \\
\hline
\multirow{3}{*}{Ala}                  & PBE+vdW   & 24 (0.5) & 44 (1.0) \\
                                      & PBE+MBD*  & 23 (0.5) & 44 (1.0) \\
                                      & PBE0+MBD* & 13 (0.3) & 28 (0.6) \\
\hline
\multirow{3}{*}{Phe}                  & PBE+vdW   & 25 (0.6) & 78 (1.8) \\
                                      & PBE+MBD*  & 26 (0.6) & 77 (1.8) \\
                                      & PBE0+MBD* & 16 (0.4) & 57 (1.3) \\
\hline
\multirow{3}{*}{Ala+Ca$^{2+}$}        & PBE+vdW   & 17 (0.4) & 23 (0.5) \\
                                      & PBE+MBD*  & 15 (0.3) & 22 (0.5) \\
                                      & PBE0+MBD* &  9 (0.2) & 15 (0.3) \\
\hline
\multirow{3}{*}{Phe+Ca$^{2+}$}        & PBE+vdW   & 105 (2.4) & 225 (5.2) \\
                                      & PBE+MBD*  &  61 (1.4) & 146 (3.4) \\
                                      & PBE0+MBD* &  50 (1.2) & 104 (2.4) \\
\hline
\end{tabular}
\label{tbl:rmsd-dfa-vs-mp2}
\end{table}


\clearpage
\noindent Further tables are provided in a Microsoft Excel file and as tab-delimited text files as Supporting Information to this article:
\begin{description}

\item[Table S1] Parameters specific to the REMD simulations of the different systems: the number of \textsl{Replicas}, the probability of \textsl{Acceptance} as well as the \textsl{Time} between exchange attempts, and the \textsl{Temperature} range of the replicas.

\item[Table S2] Number of conformers found in the different stages (after global search and after refinement) of the search scheme for amino acids, dipeptides, and complexes thereof with Ca$^{2+}$ cations. For the amino acids, the basin hopping search was performed starting from the non-zwitterionic as well as from the zwitterionic state. These numbers are separated by a ``+'' in the respective column.

\item[Table S3] Numbers of conformers found for the amino acids (AA) and their complexes with the investigated divalent cations.

\item[Table S4] Numbers of conformers found for the dipeptides (Dip.) and their complexes with the investigated divalent cations.

\item[Table S5a] Extrapolation of SCF energies as proposed by Karton and Martin: $E^{n}_{SCF} = E^{CBS}_{SCF} + A * e^{(-alpha * \sqrt{n})}$ with $n = 3,4,5$; $A$, $alpha$, $E^{CBS}_{SCF}$ to be determined by a least squares fit; perfect fit as $\# parameters = \#datapoints = 3$; all values in eV.

\item[Table S5b] Extrapolation of MP2 correlation energies as proposed by Truhlar: $E^{n}_{corr} = E^{CBS}_{corr} + B * n^{-beta}$ with $n = 3,4,5$; $B$, $beta$, $E^{CBS}_{corr}$ to be determined by a least squares fit; perfect fit as perfect fit as $\# parameters = \#datapoints = 3$; all values in eV.

\item[Table S6] Basis set superposition errors (BSSE) for SCF and MP2 correlation energies with $ n = T/Q/5$; all values in eV.

\item[Table S7] Relative gas-phase Ca$^{2+}$ binding energies for the amino acids from experiments by Ho \textit{et al.}\cite{rcms21_1097} and absolute binding energies in the gas phase from DFT-PBE+vdW calculations for amino acids and dipeptides.

\end{description}

%
%
%
%
%
%
%



\section*{Data Citations}

Bibliographic information for the data records described in the manuscript.
\begin{description}
\item[Data citation 1] Ropo, M., Baldauf, C., Blum, V. NOMAD repository. DOI: 10.17172/NOMAD/20150526220502 (2015).
\item[Data citation 2] Ropo, M., Schneider, M., Baldauf, C., Blum, V. DRYAD. DOI: 10.5061/dryad.vd177 (2016).
\end{description}


\end{document}